\documentclass[conference]{IEEEtran}
\IEEEoverridecommandlockouts
\usepackage{cite}
\usepackage{amsmath,amssymb,amsfonts}
\usepackage{algorithmic}
\usepackage{graphicx}
\usepackage{textcomp}
\usepackage{xcolor}
\usepackage{color}
\usepackage{flushend}

\def\BibTeX{{\rm B\kern-.05em{\sc i\kern-.025em b}\kern-.08em
    T\kern-.1667em\lower.7ex\hbox{E}\kern-.125emX}}
\begin{document}

\title{Wearable Sensory Substitution for \\Proprioception via Deep Pressure\\
\thanks{This work was supported in part by a grant from the RTW Charitable Foundation. SK and BBV were supported by the National Science Foundation Graduate Fellowship Program. TCB was supported by the intramural research program of the NIH Clinical Center. Work in ATC’s lab was supported by intramural funds of NCCIH and NINDS. Work in CGB’s lab was supported by intramural funds of NINDS.}
}

\author{\IEEEauthorblockN{Sreela Kodali}
\IEEEauthorblockA{\textit{Dept. of Electrical Engineering} \\
\textit{Stanford University}\\
Stanford, USA \\
kodali@stanford.edu}
\and
\IEEEauthorblockN{Brian B. Vuong}
\IEEEauthorblockA{\textit{Dept. of Mechanical Engineering} \\
\textit{Stanford University}\\
Stanford, USA \\
bbvuong@stanford.edu}
\and
\IEEEauthorblockN{Thomas C. Bulea}
\IEEEauthorblockA{\textit{Dept. of Rehabilitation Medicine} \\
\textit{National Institutes of Health}\\
Bethesda, USA \\
thomas.bulea@nih.gov}
\and
\IEEEauthorblockN{Alexander T. Chesler}
\IEEEauthorblockA{\textit{Complementary and Integrative Health} \\
\textit{National Institutes of Health}\\
Bethesda, USA \\
alexander.chesler@nih.gov}
\and
\IEEEauthorblockN{Carsten G. Bönnemann}
\IEEEauthorblockA{\textit{Neurological Disorders and Stroke} \\
\textit{National Institutes of Health}\\
Bethesda, USA \\
carsten.bonnemann@nih.gov}
\and
\IEEEauthorblockN{Allison M. Okamura}
\IEEEauthorblockA{\textit{Dept. of Mechanical Engineering} \\
\textit{Stanford University}\\
Stanford, USA \\
aokamura@stanford.edu}
}




\maketitle

\begin{abstract}
We propose a sensory substitution device that communicates one-degree-of-freedom proprioceptive feedback via deep pressure stimulation on the arm. The design is motivated by the need for a feedback modality detectable by individuals with a genetic condition known as PIEZO2 loss of function, which is characterized by absence of both proprioception and sense of light touch. We created a wearable and programmable prototype that applies up to 15 N of deep pressure stimulation to the forearm and includes an embedded force sensor. We conducted a study to evaluate the ability of participants without sensory impairment to control the position of a virtual arm to match a target angle communicated by deep pressure stimulation. A participant-specific calibration resulted in an average minimum detectable force of 0.41 N and maximum comfortable force of 6.42 N. We found that, after training, participants were able to significantly reduce angle error using the deep pressure haptic feedback compared to without it. Angle error increased only slightly with force, indicating that this sensory substitution method is a promising approach for individuals with PIEZO2 loss of function and other forms of sensory loss.
\end{abstract}

\begin{IEEEkeywords}
sensory substitution, wearable devices, proprioception, haptics
\end{IEEEkeywords}

\section{Introduction}
Proprioception is considered our ``sixth sense.'' It provides continuous information about body position and movement vital to motor control and coordination, balance, muscle tone, postural reflexes, and skeletal alignment. Many neuromuscular disorders arise from dysfunction of motor efferents, but a deficiency of afferent proprioceptive sensory input is another, often overlooked, cause of impairment that can severely impact motor function, even when strength is preserved. A large host of diseases result in loss of sensation (known as sensory neuropathies), sometimes affecting proprioception and sometimes other touch sensing modalities including vibration, skin deformation, temperature, and pain sensation. 

Proprioception in humans is entirely dependent on the non-redundant mechanosensor PIEZO2. Our long-term goal is to address lack of proprioception in individuals with recessive PIEZO2 loss of function (LOF), who show complete congenital absence of proprioception leading to motor and functional impairment \cite{piezo2_nEngJMed}. Individuals with PIEZO2-LOF also lack vibratory sense and discriminatory touch perception specifically on glabrous skin, but preserve deep pressure, temperature, and some pain sensation \cite{typeA_afferents, chesler2018piezo, piezo2_nEngJMed, szczot2018piezo2}.
\textit{Deep pressure} denotes a significant, innocuous force applied perpendicular to the skin. Applied at a low frequency, deep pressure deliberately engages both cutaneous mechanoreceptors like Merkel cells as well as type A$\beta$ sensory fibers found deeper in the tissue \cite{typeA_afferents}. No pharmacologic or assistive technology options currently exist for individuals with PIEZO2-LOF. Our goal is to design and test a wearable haptic device that enables proprioceptive feedback using preserved sensory input modalities and evaluate its efficacy for intuitive control of limb movement in individuals with PIEZO2-LOF. 

Prior work in sensory substitution for proprioception focuses on conveying the state of a prosthetic hand, arm, or leg to a user on an intact body location for an individual with amputation. For the hand, Cheng et al.\ conveyed the configuration of a virtual human hand (representing a prosthetic hand) using vibrotactile feedback on a belt worn around the waist~\cite{ChengConveying2023}. Wheeler et al.\ applied rotational skin stretch to the forearm in order to provide proprioceptive feedback from a virtual prosthetic arm controlled with myoelectric sensors \cite{WheelerInvestigation2010}. On the lower limb, Welker et al.\ controlled the position of an ankle-foot prosthesis using the wrist, effectively substituting wrist angle for ankle angle \cite{WelkerTeleoperation2021}. Skin stretch devices were used by Kayhan et al.~\cite{KayhanSkin2018} and Colella et al. \cite{ColellaNovel2019} to substitute for proprioception in multiple degrees of freedom and at multiple locations on the body. Sensory neuropathy is another condition that is often a direct complication of another comorbidity, such as stroke and diabetes.  Tzorakoleftherakis et al.\ showed that vibrotactile actuators could also convey movement of the arm for patients with loss of proprioception after stroke~\cite{TzorakoleftherakisTactile2015}. 

Unfortunately, none of the above feedback modalities are appropriate for proprioceptive feedback for individuals with PIEZO2-LOF due to their lack of cutaneous touch sensation. Of the numerous wearable haptic devices developed for arm and upper limb applications~\cite{wearablesReview2022}, a subset
has explored pressure sensations that are promising for this application. Researchers have used squeezing \cite{GuptaClench2017, PohlSqueeze2017, Patterson1992, Tejero2012}, local deep pressure \cite{Casini2015}, and combinations of modalities \cite{Aggravi2018, Dunkelberger2021}.


Because deep pressure sensation is intact for individuals with PIEZO2-LOF, we propose a sensory substitution device that communicates proprioception via deep pressure stimulation. We designed a wearable device to provide this stimulation and performed a study to determine whether participants with intact sensation could use the stimulation as a substitute for the elbow angle of a virtual arm. Our research questions are: 
\begin{itemize}
\item  What range of forces applied to the forearm are noticeable and comfortable for participants?
\item Can participants learn to map deep pressure applied to the forearm to the angles of a virtual elbow?
\item What is participants' accuracy with deep pressure stimulation, compared to when vision is used?
\item Given that sensitivity to changes in force decreases when force magnitude increases (per Weber's Law), will participants' accuracy change with force or elbow angle?
\end{itemize}
The answers to these questions will inform future designs of wearable haptic devices for sensory substitution.

\section{System Design and Experimental Setup}

\subsection{Wearable Deep Pressure Stimulation Device}
\begin{figure}[t]
\begin{center}
\includegraphics[width=8cm]{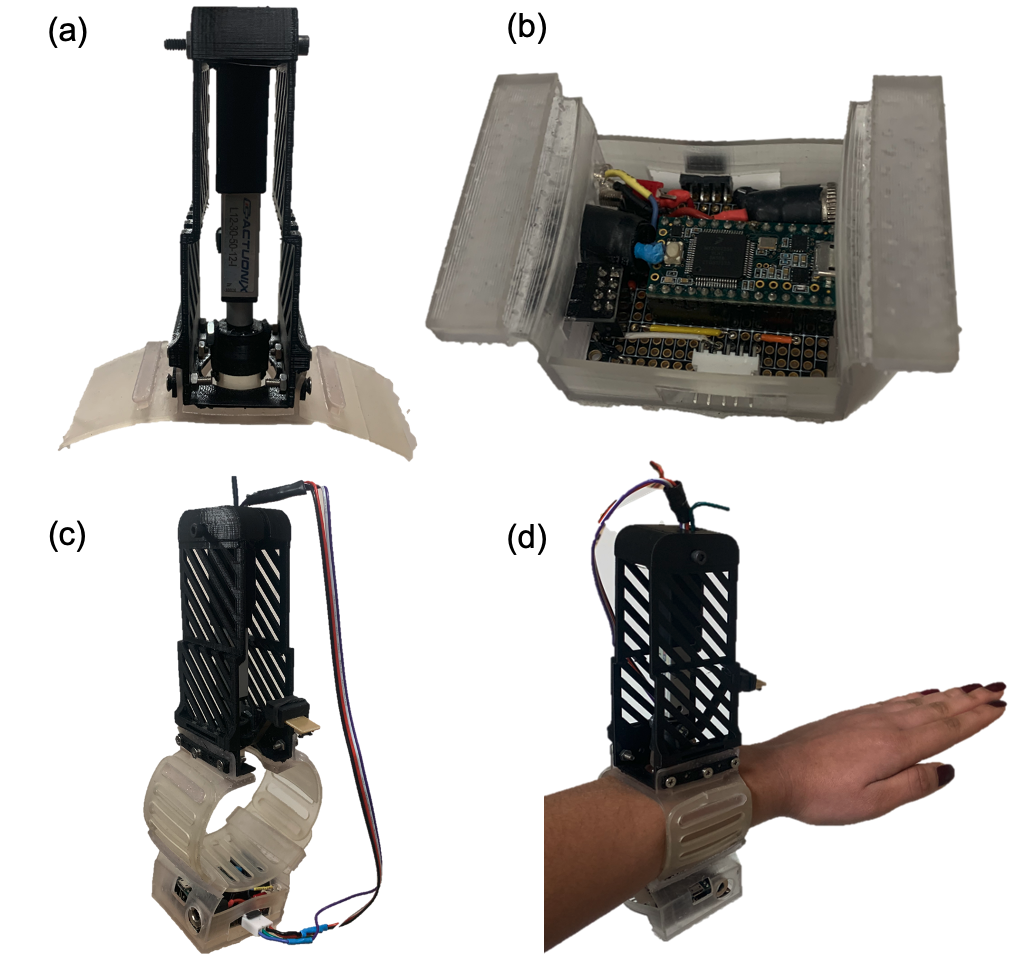}
\end{center}
\caption{The wearable device consists of (a) a linear actuator with attached force sensor and (b) embedded electronics system. The parts connect with a (c) flexible arm band and are (d) fastened to the dorsal side of the arm.}
\label{fig:device}
\end{figure}

A device was designed to communicate a user's elbow angle via deep pressure stimuli applied to the forearm. The device includes (1) a deep pressure stimulator and (2) an embedded electronics system. Deep pressure is applied by a position-controlled micro linear actuator with position feedback (Actuonix L12-30-50-12-I) and a cylindrical tactor (15 mm dia.) (Figure~\ref{fig:device}). The actuator is housed in a rigid enclosure with a flexible plastic interface for contact with skin. Plastic straps hold the device in place. As the actuator extends, the tactor presses directly on the skin and applies a deep pressure stimulus. A low-profile capacitive force sensor (SingleTact, dia.\ 15 mm and force range 45 N) was embedded into the tactor to measure the applied pressure in real time. The tactor, enclosure, interface, and straps were 3D printed. 
The stimulator weighs 96.0 g and is 49 $\times$ 38 $\times$ 120 mm.

 \begin{figure}[t]
\begin{center}
\includegraphics[width=8cm]{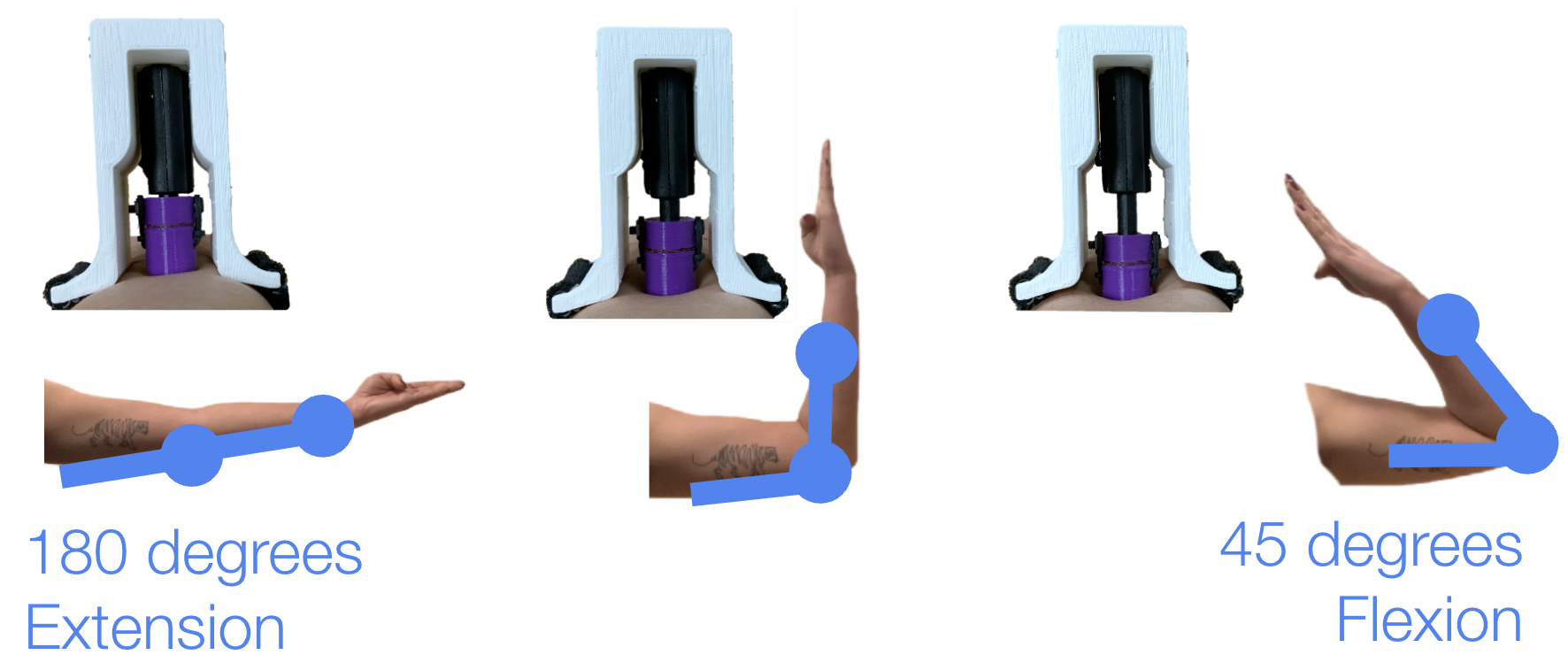}
\end{center}
\caption{With the arm fully extended at 180 degrees, the tactor is just in contact with the forearm surface and does not apply significant pressure. As the flexion angle decreases to 45 degrees (i.e., the arm becomes more flexed), the actuator moves the tactor down toward the forearm and increases the amount of deep pressure stimulation.}
\label{fig:armFlexAngles}
\end{figure}

 The embedded electronics system consists of a microcontroller (ARM Cortex-M4, Teensy 3.2), safety button, force sensor interface board,  and SD card writer. The microcontroller was programmed in C and implements a finite state machine for the device's two modes: calibration and runtime. In calibration, the microcontroller adjusts the device parameters to define the maximum and minimum positions to be used in the study. These correspond to the maximum comfortable force and minimum perceivable force selected by each user. In runtime, a local PID controller causes the actuator to move the tactor according to a target trajectory. The microcontroller records force and actuator position data. 
 The electronics box is 55 $\times$ 52 $\times$ 25 mm, weighs 57.8 g and is attached to the straps of the stimulator.

 The device is fully wearable and programmable, so the feedback mapping can be changed. The default mapping is \textbf{linear}, where deep pressure changes linearly with elbow angle (Figure~\ref{fig:armFlexAngles}). As elbow angle decreases, deep pressure increases in the same proportion.

\subsection{Virtual Environment}

\begin{figure}[t]
\centering
\includegraphics[width=8cm]{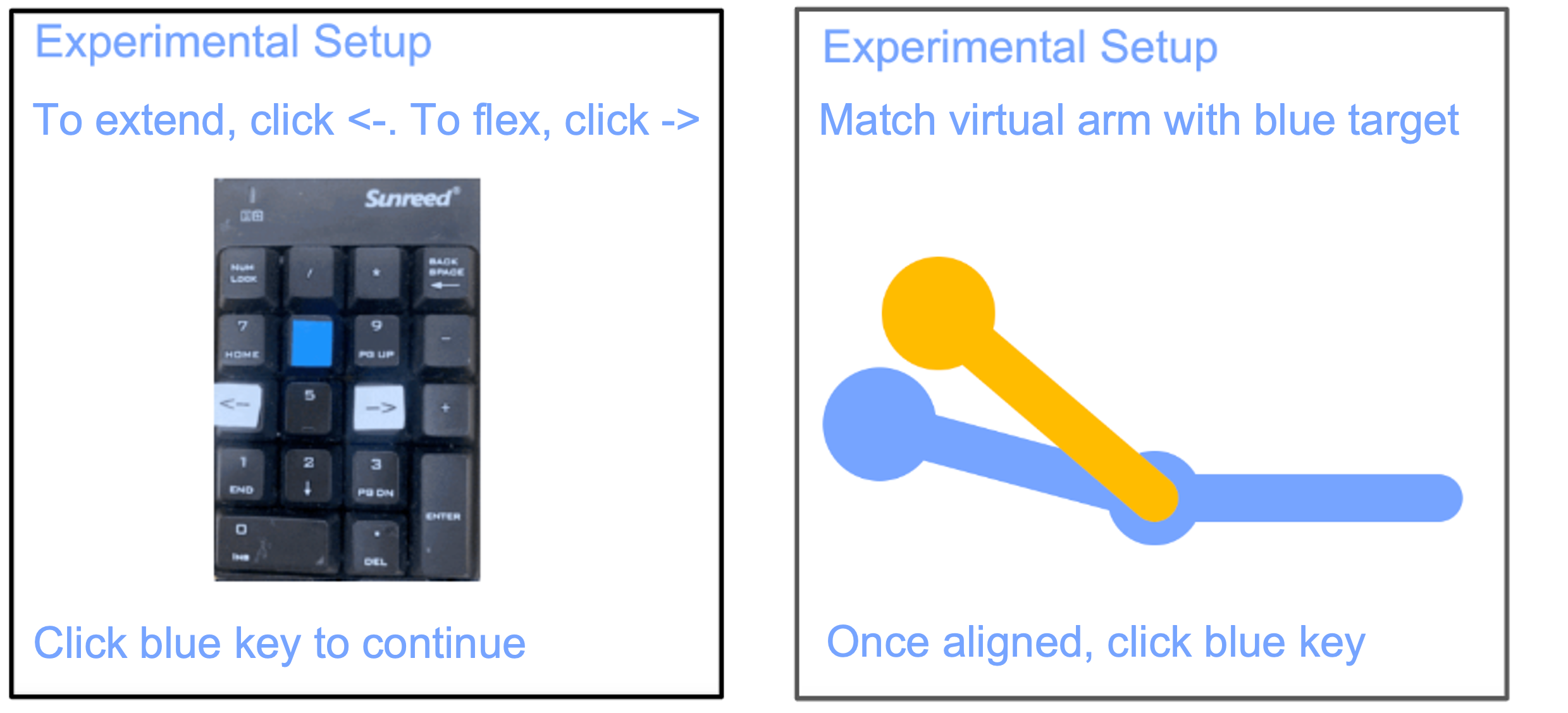}
\caption{Instructions shown on-screen for the user are: (Left) Keypad for participants to interact with the graphical user interface to adjust virtual arm with arrows and indicate completion with blue key. (Right) Example of blue target angle and yellow virtual arm graphics displayed in the virtual environment.}
\label{fig:gui}
\end{figure}

To disrupt the intact proprioception of  users without sensory impairment, we developed a custom software environment for individuals to control a virtual arm. 
A simple graphic represented the virtual arm and 1-degree-of-freedom elbow movement (Figure~\ref{fig:gui}). The left and right arrows of a keypad extend and flex the virtual arms from 180 to 45 degrees. For each key press, the virtual arm moved either 1 or 3 degrees; randomness was incorporated to prevent users from knowing the exact arm position based on the number of key presses. Users could still estimate the arm position to some extent, akin to how people use feedforward, planned motor commands to achieve motion.
As users moved the virtual arm, the arm's current position was passed to the haptic device, and the device applied the associated deep pressure stimuli.

\section{{Experimental Methods}}

\subsection{Participants}\label{AA}
 Fourteen participants were recruited through community and institution emails (ages 22-60, 7 men, 6 women, 1 declined to share gender, 12 right-handed, 2 left-handed). Individuals with diagnosed neurological or cognitive conditions were excluded from the study. The university institutional review board approved the experimental protocol, and all participants gave informed consent. 

\subsection{Procedures}

\begin{figure}[t]
\centering
\includegraphics[width=8cm]{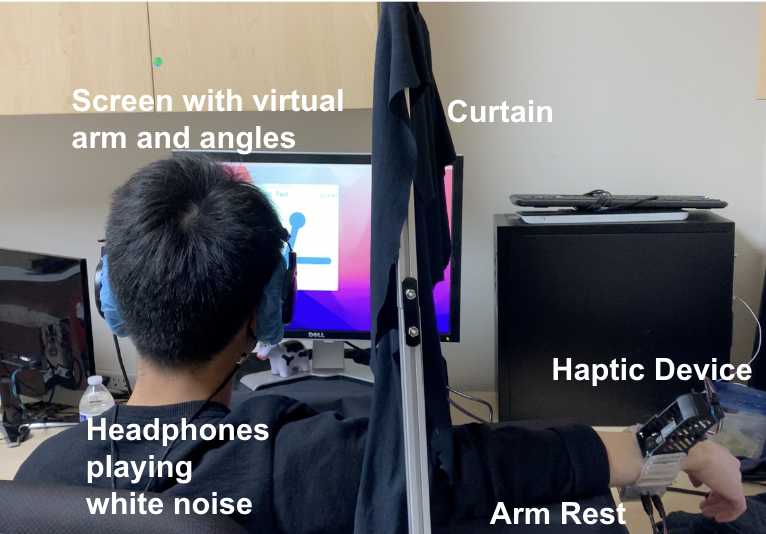}
\caption{Participant is situated in front of a display of the virtual environment, which they interact with via a keypad, while the haptic device is worn on the dominant forearm and obscured from the participant's view by a curtain.}
\label{fig:experimentalSetup}
\end{figure}

This experiment evaluates if deep pressure can be a sensory substitute for elbow angle. Participants controlled the angle of a virtual arm to match a given target angle. Target angles ranged from 180 to 45 degrees in increments of 15 degrees. 
All participants experienced conditions with and without haptic feedback to complete the task. 
Participants were randomly assigned to either test conditions with real-time haptic feedback (H) first or no haptic feedback (nH) first. To establish a baseline of optimal proprioceptive performance, participants were evaluated when they experienced visual feedback, where their virtual arm was displayed on the screen. Within these H and nH test conditions, the participants would go through one test condition with no visual feedback (nV) first and then a second test condition with real-time visual feedback (V), in that order. Each test condition consisted of a set of 10 target angles from 180 to 45 degrees, referred to as a \textit{block}.

Participants visited the lab for an approximately 1-hour long session. They started with a pre-experiment survey collecting information on prior experience with human-machine interactive devices and confirm adherence to the inclusion criteria. Participants sat in front of a computer screen displaying the virtual environment. The experimenter fastened the device to the participant's dominant arm, obtaining participant approval before touching and proceeding for each step~\cite{traumaInformed2021}. The device was positioned on the forearm's dorsal side to prevent deep pressure from interfering with blood circulation. The arm with the device rested in an extended position on an arm rest with a curtain to obscure their view of the device. Participants used their non-dominant hand to control the virtual arm with a keypad. Participants wore headphones playing white noise to avoid bias from auditory cues produced by actuator motor sounds. Each session had calibration, learning, and testing.

\subsubsection{Calibration}
After the device was securely fastened to the participant’s arm, a calibration sequence was performed to determine the minimum detection pressure and maximum comfortable deep pressure. For minimum pressure, the actuator started from a position not contacting the forearm surface and gradually moved toward the forearm surface. The participant verbally notified the experimenter when they first felt pressure, and the corresponding actuator position and force measured were recorded. For maximum pressure, the actuator gradually moved while applying pressure on the forearm, and the participants verbally noted when they would start feeling uncomfortable. This calibration process was a dialogue between participant and experimenter and was repeated at least three times per participant. The calibrated actuator positions for maximum and minimum pressures were stored and linearly mapped to the minimum and maximum virtual arm angles.

\subsubsection{Learning}
Learning was a robust sequence of tasks to ensure the participant had adequate practice with deep pressure feedback. Learning consisted of four phases:
\begin{itemize}
    \item \textit{Explore:} Participants moved their virtual arm freely, saw their virtual arm move on the display, observed the associated haptic feedback in real time, and learned to associate deep pressure haptic feedback with the virtual arm’s angle. This phase lasted for 1 minute.
    \item \textit{Target:} Participants moved their virtual arm, which was displayed in real time on the screen, to target angles and were instructed to pay attention to the haptic feedback once they reached the target angles. Participants had unlimited time for each angle and went through two blocks (20 angles total). The first block was in descending order, and second block was in random order.
    \item \textit{Haptic Feedback:} Participants passively experienced the haptic feedback associated with each target angle in descending order, for a minimum duration of 10 seconds and for as long as the participant needed. Participants did not have a virtual arm to move in this phase.
    \item \textit{Practice:} Participants matched their virtual arm, which was not shown on the screen, to a displayed target angle using only real-time haptic feedback. For each target angle, once the participant indicated the completion of their angle-matching attempt with a key press, their virtual arm's current position would appear on screen with the target angle for 10 seconds to provide corrective feedback so participants could learn. Practice included four blocks, each with differing orders (40 angles total). Block 1 was in descending order. Block 2 had part ascending, part descending, and a few angles interspersed randomly. Blocks 3 and 4 were in random order.
\end{itemize}
At the start of each of these learning phases, the virtual arm's position was reset to 180 degrees. The virtual arm’s position was not reset between angles within a learning phase. At the end of a learning phase, no haptic feedback was applied while participants received instructions for the next section.

\subsubsection{Testing}
Upon completion of learning and an optional break, participants proceeded to testing. For each of the four test conditions (H nV; H V; nH nV; and nH V), participants received one block of target angles in random order and tried to match the virtual arm position to the displayed target angle.

After testing, participants completed a post-experiment survey that asked participants to rate the level of difficulty in completing the angle-matching tasks for each of the four test conditions and describe their strategies. 

\subsection{Metrics}
We quantified proprioceptive performance by the difference between the virtual arm angle and the target angle, henceforth called \textit{angle error}. In our subsequent analyses, our primary metric was the absolute value of angle error, called \textit{angle error magnitude} ($|$angle error$|$), because it accounts for the full magnitude of overshooting and undershooting. Angle error that includes the sign of the error is also reported. Statistical tests were performed with R software~\cite{rSoftware}, and figures were generated with MATLAB~\cite{matlab}. We also recorded force and task completion time/speed.

\section{Results and Discussion}

\subsection{Calibration}
\begin{figure}[t]
\includegraphics[width=\columnwidth]{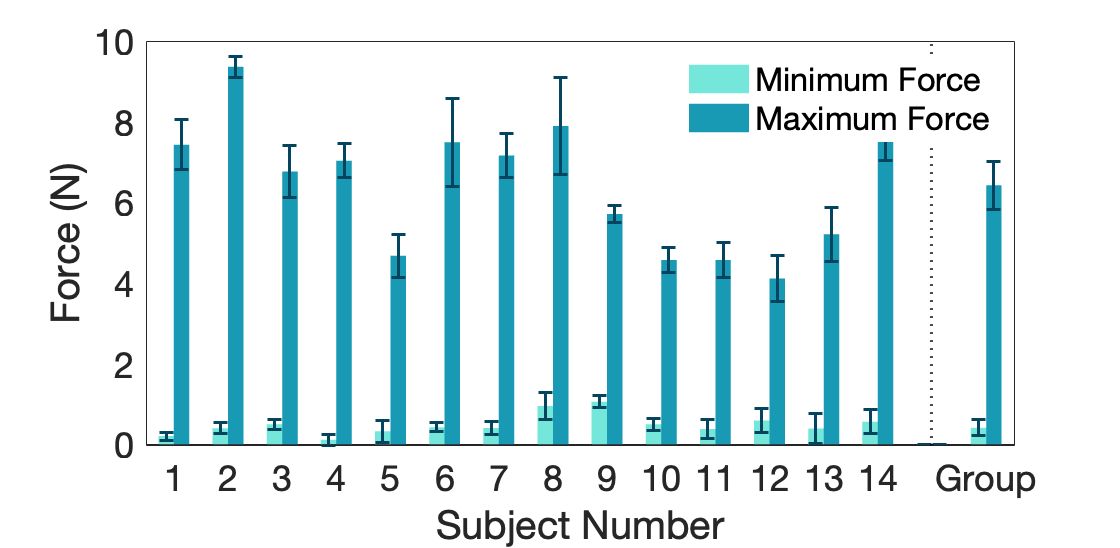}
\caption{Average and standard error of forces measured during calibration of minimum and maximum actuator positions for each subject. Across all fourteen subjects, the average minimum detection force was 0.41 N and maximum comfortable force was 6.42 N.}
\label{fig:calibration}
\end{figure}

Figure~\ref{fig:calibration} shows the minimum and maximum calibration forces selected by each participant and the group mean. Across all fourteen subjects, the mean calibration minimum and maximum forces were 0.41 N and 6.42 N, respectively.

\subsection{Angle Error}

\begin{figure}[t]
\centering
\includegraphics[width=8cm]{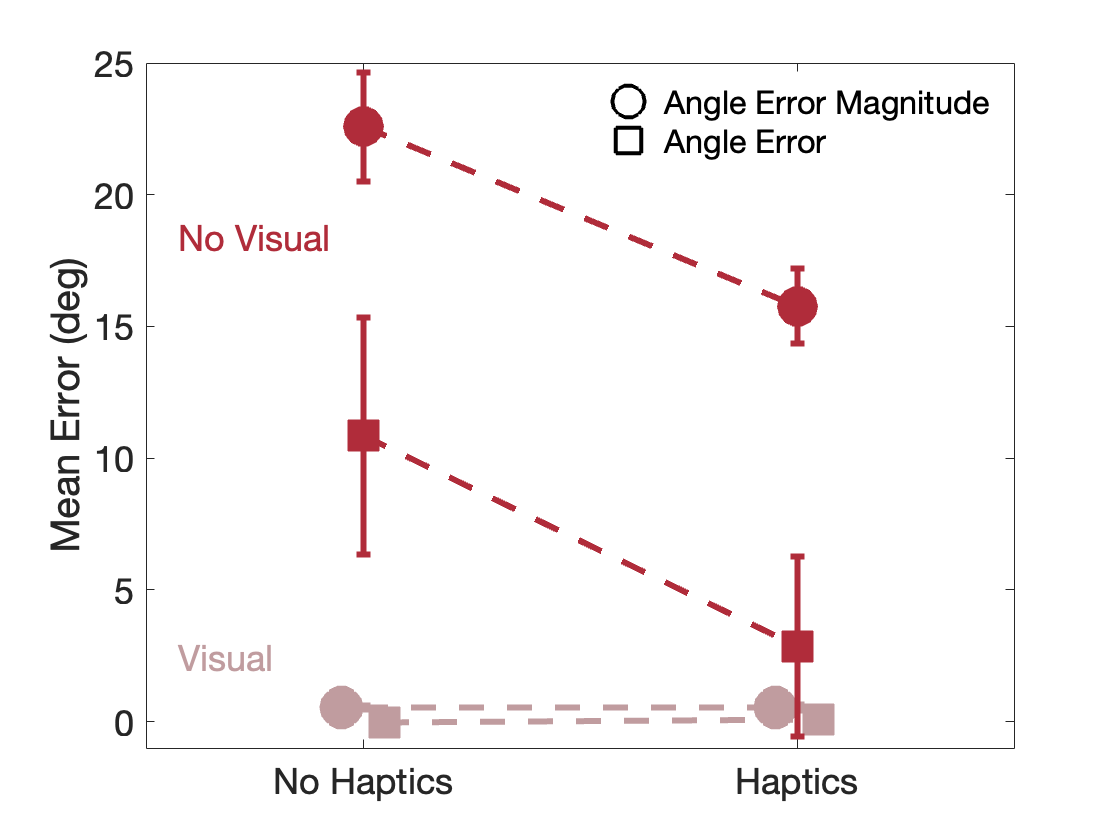} 
\caption{The mean and standard error of angle error magnitude and angle error are compared across the four conditions. When visual feedback is present, the error is minimal.  Without visual feedback, haptic feedback results in significantly less error than without haptic feedback. The mean angle error magnitudes for the conditions nH V, H V, nH nV, H nV are 0.53\textdegree, 0.54\textdegree, 22.57\textdegree, and 15.77\textdegree, respectively. The mean angle errors for the conditions nH V, H V, nH nV, H nV are 0.05\textdegree, 0.1\textdegree, 10.84\textdegree, and 2.83\textdegree, respectively.}
\label{fig:mainresult}
\end{figure}

The key takeaway of this study is that participants were able to use deep pressure feedback to improve their proprioception without vision. Figure~\ref{fig:mainresult} shows mean angle error magnitude and mean angle error for all conditions. When there is no visual display of the current position of the virtual arm, such that participant relied only on haptic feedback, the haptic feedback condition (H nV) had both a lower angle error magnitude (15.77\textdegree $\pm$ 1.42\textdegree) and angle error (2.83\textdegree $\pm$ 3.29\textdegree) than the no haptic feedback condition (nH nV)  angle error magnitude (22.57\textdegree $\pm$ 2.06\textdegree) and angle error (10.84\textdegree $\pm$ 4.33\textdegree).

Angle error magnitude is larger than angle error because both overshooting and undershooting add to the magnitude. Angle error has a larger variance because overshooting and undershooting are signed. Interestingly, mean angle error was positive, indicating that participants often overshot. The errors for no haptic and no visual feedback (nH nV) are not as large as anticipated because participants could, to some extent, estimate the number of key presses required to move the virtual arm to a target. This is akin to the use of feedforward (as opposed to feedback) control.

Both angle error magnitude and angle error without visual feedback (nV) are much larger than with visual feedback present (V). The angle error magnitudes in the visual feedback conditions are non-zero (nH V = 0.53\textdegree $\pm$ 0.09\textdegree, H V = 0.54\textdegree $\pm$ 0.12\textdegree) because, 
when movement occurred in 3 degree increments, the virtual arm often could not align exactly with the target angle. 
In contrast, angle error is nearly zero (nH V = 0.05\textdegree $\pm$ 0.07\textdegree, H V = 0.01\textdegree $\pm$ 0.06\textdegree) because the overshooting and undershooting cancel each other out. The visual feedback conditions represent the best possible performance for the experimental setup.

The subsequent analyses focus solely on angle error magnitude. A two-way repeated measures ANOVA was used to identify the main effects of visual feedback, haptic feedback, and the interaction between these two feedback types for all subjects. The results revealed main effects of haptic feedback (\(F(1, 13) = 6.87\), \(p = 0.021\), \(\eta^2 = 0.125\)) and visual feedback (\(F(1,13) = 239.67\), \(p = 9.34 \times 10^{-10}\),  \(\eta^2 = 0.810\)). The main effects were qualified by interactions between visual and haptic feedback (\(F(1,13) = 6.95\), \(p = 0.021\),  \(\eta^2 = 0.124\)).


Grouping the test data by whether or not visual feedback was provided, we performed a one-way ANOVA with haptic feedback as the within-subjects factor variable and angle error magnitude as the dependent variable. Within the no visual feedback (nV) test conditions, we observed a statistically significant effect of haptic feedback when no visual feedback was provided (\(F(1, 13) = 6.92\), \(p = 0.021\), \(\eta^2 = 0.22\)) but no such effect in the conditions with visual feedback (\(F(1, 13) = 0.02\), \(p = 0.898\), \(\eta^2 = 3.65 \times 10^4\)). The ANOVA was followed by post-hoc t-tests with Bonferroni corrections. We conducted a pairwise comparison between haptic (H) and no haptic (nH) feedback test conditions for both visual (V) and no visual (nV) feedback groups. Similarly, this analysis showed that haptic feedback had a significant effect (adjusted $p = 0.021$) on angle error magnitude in the no visual feedback (nV) conditions but not in the visual feedback (V) conditions.

\begin{figure}[t]
\centering
\includegraphics[width=8cm]{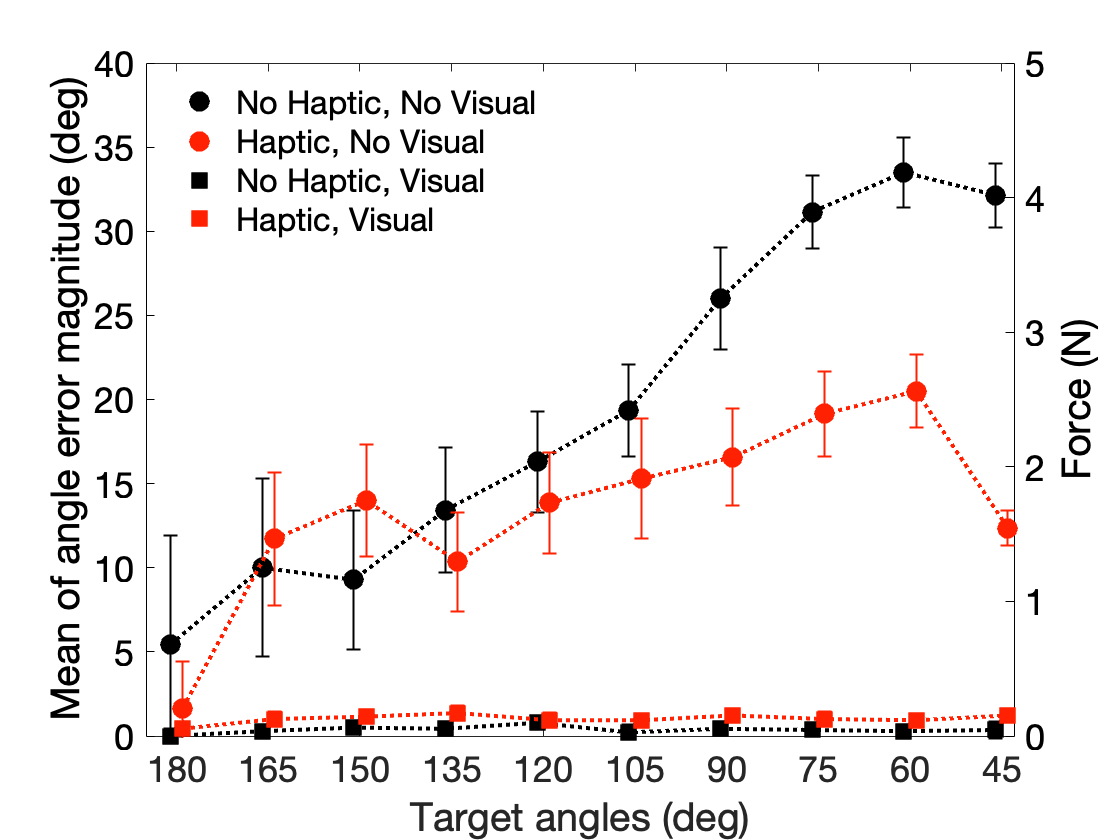}
\caption{Angle error magnitude versus target angle. For the cases with visual feedback (square data points), the angle error magnitude is slightly above zero due to the randomness we injected into the angle increment with each key press. For the case without vision and without haptic feedback (black circular data points), the angle error magnitude visibly increases with target angle as the arm becomes more flexed. For the case without vision and with haptic feedback (red circular data points), the angle error magnitude does not consistently increase with target angle as the arm becomes more flexed. The 180 degree target is the virtual arm fully extended, and the 45 degree target represents the maximum flex of the virtual arm in the study.}
\label{fig:angleerror}
\end{figure}

Figure~\ref{fig:angleerror} shows the angle error magnitude as a function of target angle. In the testing conditions with no visual feedback (nH nV and H nV), we observed an overall decrease in mean angle error magnitude as the target angle increased. In the 45 to 135 degree target angle range, angle error magnitude was consistently higher in the nH V condition than in the H V condition. The relative flattening of the error magnitude in the H nV condition indicates that participants' ability to achieve a desired target angle did not change substantially with target angle. This was somewhat surprising, given that sensitivity to changes in force decreases when force magnitude increases (per Weber's Law). We believe that the extensive training provided and the nature of the task (to achieve a target rather than perform a two-alternative forced-choice task) enabled this result. 
In conditions without visual feedback (nH nV and H nV), the decrease in error magnitude at the maximum and minimum angles can be attributed to the system design. The virtual arm’s movement was bounded from 180 to 45 degrees, and the device had a minimum and maximum pressure. Participants could use repeated key presses and the haptic stimuli bounds to reach the maximum and minimum angles, and this was corroborated by the post-experiment survey. Additionally, we found no relationship between a participant's force range (maximum comfortable pressure - minimum perceivable pressure) and task performance.

\subsection{Actuator Force}
\begin{figure}[t]
\centering
\includegraphics[width=8cm]{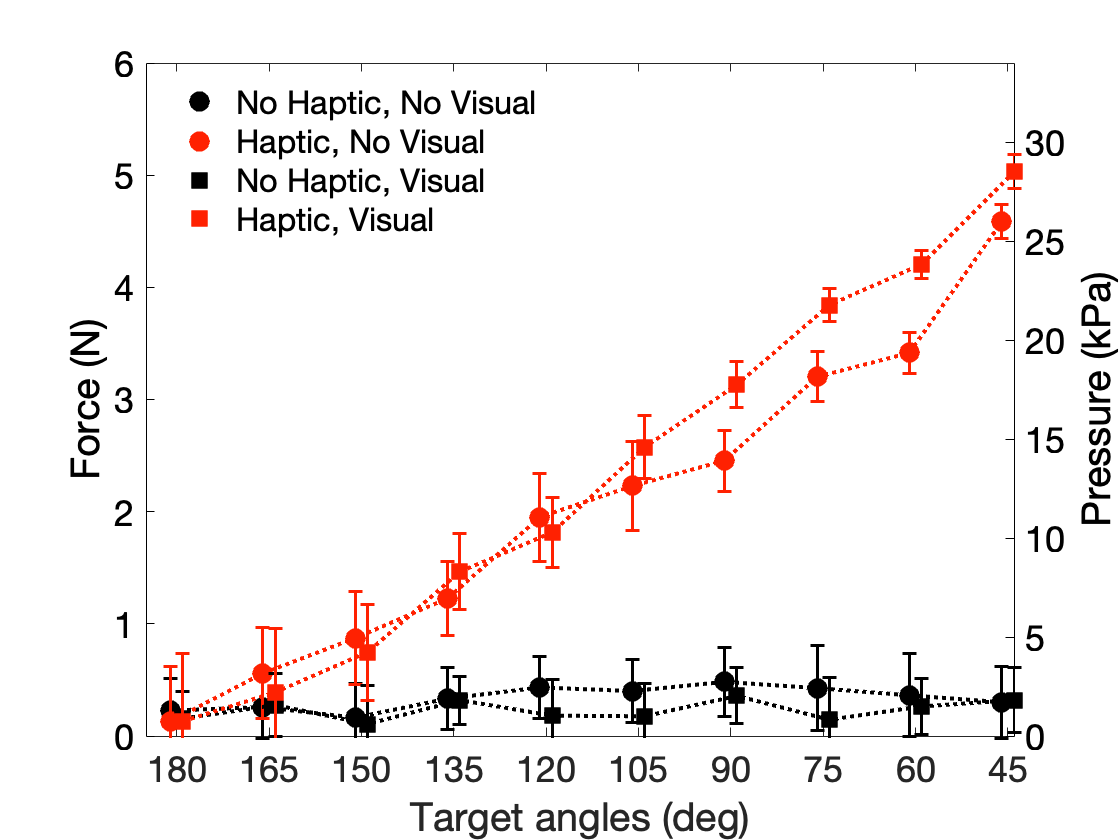}
\caption{Applied force versus target angle for the four conditions, averaged over all trials for all users. For the cases with no haptics (black data), the force is slightly above zero due to sensor noise and potentially light contact between the tactor and the skin. For the cases with haptic feedback (red data), the forces increased approximately linearly with angle as the arm becomes more flexed. The 180 degree target is the virtual arm fully extended, and the 45 degree target represents the maximum flex of the virtual arm in the study.}
\label{fig:angleVsForce}
\end{figure}

In the testing conditions with haptic feedback (H nV and H V), a roughly linear relationship between target angle and measured actuator force was observed Figure~\ref{fig:angleVsForce}. In the testing conditions without haptic feedback (nH nV and nH V), measured actuator force consistently stayed below 1 N as expected. This confirmed that our mapping from virtual arm angle to actuator position was functioning as intended.

\begin{figure}[t]
\includegraphics[width=8cm]{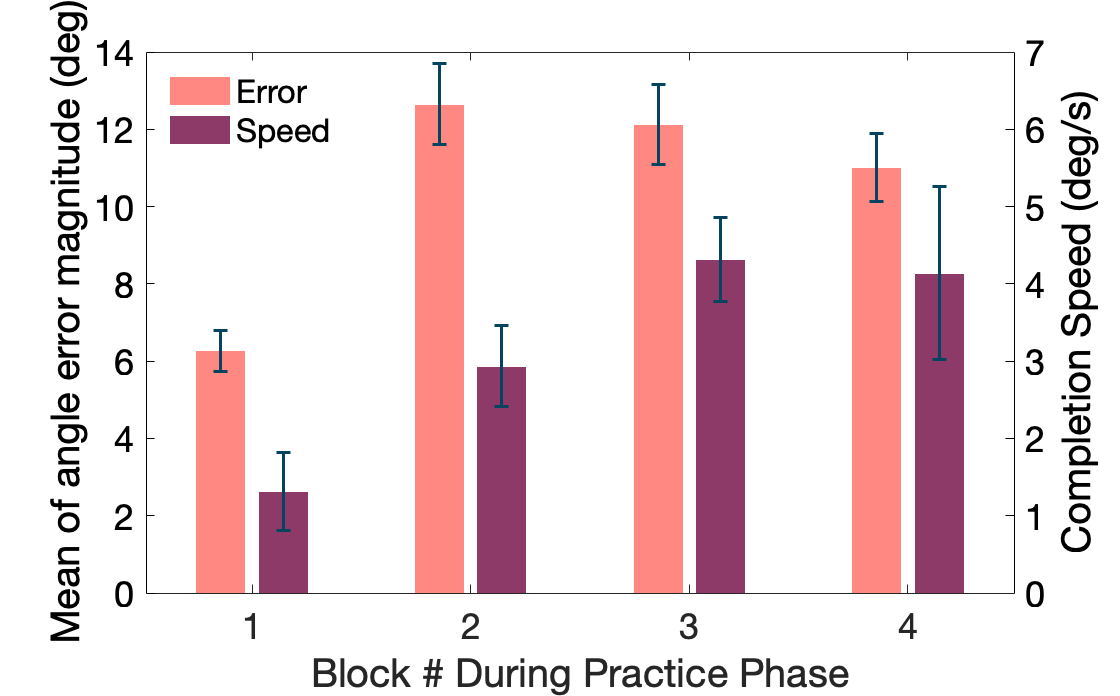}
\caption{Angle error magnitude and completion speed versus the block number in the \textit{Practice} phase. \textit{Practice} has 4 blocks of trials. In block 1, the target angles were in descending order, so angle error magnitude (orange) is low. In block 2, the angle order has randomness so angle error magnitude drastically increases. In blocks 3 and 4 where the angles are in random order, angle error magnitude reduces, confirming practice helps. For completion speed (magenta), participants begin slowly in block 1 and progressively get faster in blocks 2 and 3. The reduction in speed in block 4 indicates a speed-accuracy trade-off, where participants spent more time and perform better at the task.}
\label{fig:learning}
\end{figure}

\subsection{Learning}
Participants generally decreased their error over the course of learning. Figure~\ref{fig:learning} shows participants’ angle error magnitude and task completion speed for the four blocks of the \textit{Practice} phase of learning. In block 1, the error was low, likely because the angles were presented in descending order. For blocks 2-4, where randomness was introduced, angle error magnitude decreases across the blocks. Interestingly, participants completed the task faster with each block except for the final block. These results suggest a change in speed-accuracy trade-off during learning, with the caveat that participants were not asked to complete the task quickly. 

\subsection{Survey Results}
Participants rated the difficulty of the task for each condition on a scale of 1 to 5. The values 1-5 corresponded to the following: ``very easy", ``easy", ``moderate", ``hard", and ``very hard." Participants' difficulty ratings mirrored their task performance. With visual feedback, participants found the task to be ``very easy", with haptic feedback making it even easier (mean difficulty rating H V = 1.14, nH V = 1.29). In conditions without visual feedback, participants found the task to be much more challenging but haptic feedback helped (nH nV = 4.86, H nV = 3.43).

The survey also inquired if any strategies were employed to complete the tasks. Many participants attempted to count key presses and used the bounds of haptic feedback. A few participants stated that they paid attention to the differential and ``relative" pressure instead of the absolute pressure. 

\section{Conclusions and Future Work}

We found that participants were able to learn a mapping between haptic feedback through deep pressure stimulation and target angles of a virtual arm. The use of a virtual arm was key because our participants had intact proprioceptive sensation, and we seek to understand whether this form of haptic feedback will be effective in individuals with PIEZO2-LOF, who do not have proprioception.

We now review the answers to our original research questions. The forces applied to the forearm that were noticeable and comfortable for participants was, on average, 0.41 N for the minimum noticeable force and 6.42 N for the maximum comfortable force. 
After an extensive training session, participants learned a mapping such that their performance with haptic feedback was statistically significantly better than without haptic feedback. Errors in the condition with haptic and without visual feedback were significantly less than those in the condition without haptic and without visual feedback. This error was also significantly greater than the gold standard of near zero degrees when vision is available. Participants' accuracy with haptic feedback did change somewhat with force, but not to the extent that might be predicted by Weber's law (linearly decreasing sensitivity with stimulus intensity).

Our results indicate that deep pressure stimulation could be used a method for sensory substitution for individuals with PIEZO2-LOF, and possibly other scenarios such as amputation and sensory neuropathy. However, in our current approach, the lower range of forces used by the healthy participants in our study will not be perceptible for future participants with PIEZO2-LOF. This will decrease the range and likely the resolution with which forces can be displayed. Multiple tactors that effectively add to the force range may be required.

Our long-term engineering challenge is to generate an appropriate substituted feedback signal for individuals with PIEZO2-LOF that is effective (i.e., replaces the missing proprioceptive sensation with an intact sensation in a manner that is useful for activities of daily living) and convenient (e.g. low-cost and wearable). Thus, our next steps in the design of the system is to develop a wearable elbow angle sensor and a soft pneumatic wearable haptic device that offers a smaller size and lower weight of the worn device. Eventually, we aim to develop multi-degree-of-freedom sensory prostheses to enable improvement in the performance of functional tasks for individuals with sensory loss.

Our long-term clinical challenge is to understand intact sensory abilities in individuals and populations with proprioceptive loss, as well as their ability to learn to use a sensory prosthesis. We aim to test in individuals with PIEZO2-LOF. 
In addition to improving individuals' health and quality of life, our work aims to provide neuroscientific insight into the role of proprioception in human motor control and how humans adapt to new sensorimotor scenarios.


%

\bibliographystyle{IEEEtran}
\bibliography{myBib}

\begin{thebibliography}{10}
\providecommand{\url}[1]{#1}
\csname url@samestyle\endcsname
\providecommand{\newblock}{\relax}
\providecommand{\bibinfo}[2]{#2}
\providecommand{\BIBentrySTDinterwordspacing}{\spaceskip=0pt\relax}
\providecommand{\BIBentryALTinterwordstretchfactor}{4}
\providecommand{\BIBentryALTinterwordspacing}{\spaceskip=\fontdimen2\font plus
\BIBentryALTinterwordstretchfactor\fontdimen3\font minus
  \fontdimen4\font\relax}
\providecommand{\BIBforeignlanguage}[2]{{%
\expandafter\ifx\csname l@#1\endcsname\relax
\typeout{** WARNING: IEEEtran.bst: No hyphenation pattern has been}%
\typeout{** loaded for the language `#1'. Using the pattern for}%
\typeout{** the default language instead.}%
\else
\language=\csname l@#1\endcsname
\fi
#2}}
\providecommand{\BIBdecl}{\relax}
\BIBdecl

\bibitem{piezo2_nEngJMed}
\BIBentryALTinterwordspacing
A.~T. Chesler, M.~Szczot, D.~Bharucha-Goebel, M.~Čeko, S.~Donkervoort,
  C.~Laubacher, L.~H. Hayes, K.~Alter, C.~Zampieri, C.~Stanley, A.~M. Innes,
  J.~K. Mah, C.~M. Grosmann, N.~Bradley, D.~Nguyen, A.~R. Foley, C.~E.
  Le~Pichon, and C.~G. Bönnemann, ``The role of piezo2 in human
  mechanosensation,'' \emph{New England Journal of Medicine}, vol. 375, no.~14,
  pp. 1355--1364, 2016, pMID: 27653382. [Online]. Available:
  \url{https://doi.org/10.1056/NEJMoa1602812}
\BIBentrySTDinterwordspacing

\bibitem{typeA_afferents}
L.~K. Case, J.~Liljencrantz, N.~Madian, A.~Necaise, J.~Tubbs, M.~McCall, M.~L.
  Bradson, M.~Szczot, M.~H. Pitcher, N.~Ghitani \emph{et~al.}, ``Innocuous
  pressure sensation requires a-type afferents but not functional piezo2
  channels in humans,'' \emph{Nature communications}, vol.~12, no.~1, pp.
  1--10, 2021.

\bibitem{chesler2018piezo}
A.~T. Chesler and M.~Szczot, ``Piezo ion channels: portraits of a pressure
  sensor,'' \emph{Elife}, vol.~7, p. e34396, 2018.

\bibitem{szczot2018piezo2}
M.~Szczot, J.~Liljencrantz, N.~Ghitani, A.~Barik, R.~Lam, J.~H. Thompson,
  D.~Bharucha-Goebel, D.~Saade, A.~Necaise, S.~Donkervoort \emph{et~al.},
  ``Piezo2 mediates injury-induced tactile pain in mice and humans,''
  \emph{Science translational medicine}, vol.~10, no. 462, p. eaat9892, 2018.

\bibitem{ChengConveying2023}
A.~Cheng, K.~A. Nichols, H.~M. Weeks, N.~Gurari, and A.~M. Okamura, ``Conveying
  the configuration of a virtual human hand using vibrotactile feedback,'' in
  \emph{2012 IEEE Haptics Symposium (HAPTICS)}, 2012, pp. 155--162.

\bibitem{WheelerInvestigation2010}
J.~Wheeler, K.~Bark, J.~Savall, and M.~Cutkosky, ``Investigation of rotational
  skin stretch for proprioceptive feedback with application to myoelectric
  systems,'' \emph{IEEE Transactions on Neural Systems and Rehabilitation
  Engineering}, vol.~18, no.~1, pp. 58--66, 2010.

\bibitem{WelkerTeleoperation2021}
C.~G. Welker, V.~L. Chiu, A.~S. Voloshina, S.~H. Collins, and A.~M. Okamura,
  ``Teleoperation of an ankle-foot prosthesis with a wrist exoskeleton,''
  \emph{IEEE Transactions on Biomedical Engineering}, vol.~68, no.~5, pp.
  1714--1725, 2021.

\bibitem{KayhanSkin2018}
O.~Kayhan, A.~K. Nennioglu, and E.~Samur, ``A skin stretch tactor for sensory
  substitution of wrist proprioception,'' in \emph{2018 IEEE Haptics Symposium
  (HAPTICS)}, 2018, pp. 26--31.

\bibitem{ColellaNovel2019}
N.~Colella, M.~Bianchi, G.~Grioli, A.~Bicchi, and M.~G. Catalano, ``A novel
  skin-stretch haptic device for intuitive control of robotic prostheses and
  avatars,'' \emph{IEEE Robotics and Automation Letters}, vol.~4, no.~2, pp.
  1572--1579, 2019.

\bibitem{TzorakoleftherakisTactile2015}
E.~Tzorakoleftherakis, M.~C. Bengtson, F.~A. Mussa-Ivaldi, R.~A. Scheidt, and
  T.~D. Murphey, ``Tactile proprioceptive input in robotic rehabilitation after
  stroke,'' in \emph{2015 IEEE International Conference on Robotics and
  Automation}, 2015, pp. 6475--6481.

\bibitem{wearablesReview2022}
A.~Adilkhanov, M.~Rubagotti, and Z.~Kappassov, ``Haptic devices:
  Wearability-based taxonomy and literature review,'' \emph{IEEE Access},
  vol.~10, pp. 91\,923--91\,947, 2022.

\bibitem{GuptaClench2017}
\BIBentryALTinterwordspacing
A.~Gupta, A.~A.~R. Irudayaraj, and R.~Balakrishnan, ``Hapticclench:
  Investigating squeeze sensations using memory alloys,'' in \emph{Proceedings
  of the 30th Annual ACM Symposium on User Interface Software and Technology},
  ser. UIST '17.\hskip 1em plus 0.5em minus 0.4em\relax New York, NY, USA:
  Association for Computing Machinery, 2017, p. 109–117. [Online]. Available:
  \url{https://doi.org/10.1145/3126594.3126598}
\BIBentrySTDinterwordspacing

\bibitem{PohlSqueeze2017}
\BIBentryALTinterwordspacing
H.~Pohl, P.~Brandes, H.~Ngo~Quang, and M.~Rohs, ``Squeezeback: Pneumatic
  compression for notifications,'' in \emph{Proceedings of the 2017 CHI
  Conference on Human Factors in Computing Systems}, ser. CHI '17.\hskip 1em
  plus 0.5em minus 0.4em\relax New York, NY, USA: Association for Computing
  Machinery, 2017, p. 5318–5330. [Online]. Available:
  \url{https://doi.org/10.1145/3025453.3025526}
\BIBentrySTDinterwordspacing

\bibitem{Patterson1992}
P.~Patterson and J.~Katz, ``Design and evaluation of a sensory feedback system
  that provides grasping pressure in a myoelectric hand,'' \emph{Journal of
  rehabilitation research and development}, vol.~29, pp. 1--8, 02 1992.

\bibitem{Tejero2012}
C.~Tejeiro, C.~E. Stepp, M.~Malhotra, E.~Rombokas, and Y.~Matsuoka,
  ``Comparison of remote pressure and vibrotactile feedback for prosthetic hand
  control,'' in \emph{IEEE International Conference on Biomedical Robotics and
  Biomechatronics (BioRob)}, 2012, pp. 521--525.

\bibitem{Casini2015}
S.~Casini, M.~Morvidoni, M.~Bianchi, M.~Catalano, G.~Grioli, and A.~Bicchi,
  ``Design and realization of the cuff - clenching upper-limb force feedback
  wearable device for distributed mechano-tactile stimulation of normal and
  tangential skin forces,'' in \emph{2015 IEEE/RSJ International Conference on
  Intelligent Robots and Systems (IROS)}, 2015, pp. 1186--1193.

\bibitem{Aggravi2018}
M.~Aggravi, F.~Pausé, P.~R. Giordano, and C.~Pacchierotti, ``Design and
  evaluation of a wearable haptic device for skin stretch, pressure, and
  vibrotactile stimuli,'' \emph{IEEE Robotics and Automation Letters}, vol.~3,
  no.~3, pp. 2166--2173, 2018.

\bibitem{Dunkelberger2021}
N.~Dunkelberger, J.~L. Sullivan, J.~Bradley, I.~Manickam, G.~Dasarathy,
  R.~Baraniuk, and M.~K. O’Malley, ``A multisensory approach to present
  phonemes as language through a wearable haptic device,'' \emph{IEEE
  Transactions on Haptics}, vol.~14, no.~1, pp. 188--199, 2021.

\bibitem{traumaInformed2021}
\BIBentryALTinterwordspacing
B.~eth Israel Deaconness Medical Center Center~for Violence~Prevention and
  R.~Staff. (2021, Jun) Trauma informed care tips. [Online]. Available:
  \url{https://www.bidmc.org/-/media/files/beth-israel-org/centers-and-departments/social-work/center-for-violence-prevention-and-recovery/bidmc-cvpr-trauma-informed-care-tips-sheet-062921.pdf}
\BIBentrySTDinterwordspacing

\bibitem{rSoftware}
\BIBentryALTinterwordspacing
{R Core Team}, \emph{R: A Language and Environment for Statistical Computing},
  R Foundation for Statistical Computing, Vienna, Austria, 2021. [Online].
  Available: \url{https://www.R-project.org/}
\BIBentrySTDinterwordspacing

\bibitem{matlab}
\BIBentryALTinterwordspacing
T.~M. Inc., \emph{MATLAB version: 9.13.0 (R2022b)}, Natick, Massachusetts,
  United States, 2022. [Online]. Available: \url{https://www.mathworks.com}
\BIBentrySTDinterwordspacing

\end{thebibliography}

\end{document}